\PassOptionsToPackage{table, dvipsnames}{xcolor}

\documentclass[sigconf, authorversion, noacm]{acmart}

\usepackage{mystyle}

\togglefalse{highlight-revision}

\def\doctitle{On-the-Fly Syntax Highlighting using Neural Networks}

\def\docauthors{Marco Edoardo Palma, Pasquale Salza, Harald C. Gall}
\def\dockeywords{%
Syntax highlighting, neural networks, deep learning, regular expressions
}

\StrSubstitute{\doctitle}{\\}{ }[\cleandoctitle]
\StrSubstitute{\dockeywords}{.}{}[\cleandockeywords]
\hypersetup{
	pdftitle={\cleandoctitle},
	pdfauthor={\docauthors},
	pdfkeywords={\cleandockeywords}
}

\def\replicationurl{https://hlnn.netlify.app}

\DeclareAcronym{api}{
	short = API,
	long = {Application Program Interface}
}

\DeclareAcronym{bert}{
	short = BERT,
	long = {Bidirectional Encoder Representations from Transformers}
}

\DeclareAcronym{ast}{
	short = AST,
	long = {Abstract Syntax Tree}
}

\DeclareAcronym{cfg}{
	short = CFG,
	long = {Control Flow Graph}
}

\DeclareAcronym{gpt}{
	short = GPT,
	long = {Generative Pretrained Transformer}
}

\DeclareAcronym{ir}{
	short = IR,
	long = {Information Retrieval}
}

\DeclareAcronym{lstm}{
	short = LSTM,
	long = {Long Short-Term Memory}
}

\DeclareAcronym{nn}{
	short = NN,
	long = {Neural Network}
}

\DeclareAcronym{rnn}{
	short = RNN,
	long = {Recurrent Neural Network}
}

\DeclareAcronym{brnn}{
	short = BRNN,
	long = {Bidirectional Recurrent Neural Network}
}

\DeclareAcronym{cnn}{
	short = CNN,
	long = {Convolutional Neural Network}
}

\DeclareAcronym{tf-idf}{
	short = tf-idf,
	long = {term frequency–-inverse document frequency}
}

\DeclareAcronym{anova}{
	short = ANOVA,
	long = {ANalysis Of VAriance}
}

\DeclareAcronym{ta}{
	short = TA,
	long = {Task-Adaptive}
}

\DeclareAcronym{gru}{
	short = GRU,
	long = {Gated Recurrent Unit}
}

\DeclareAcronym{sh}{
	short = SH,
	long = {syntax highlighting}
}

\DeclareAcronym{re}{
	short = regex,
	short-plural = es,
	long = {regular expression}
}

\DeclareAcronym{om}{
	short = OM,
	long = {Oracle Methods}
}

\DeclareAcronym{eta}{
	short = ETA,
	long = {Extended Token Annotation}
}

\DeclareAcronym{heta}{
	short = HETA,
	long = {Highlighted Extended Token Annotation}
}

\DeclareAcronym{ide}{
	short = IDE,
	long = {Integrated Development Environment}
}

\DeclareAcronym{bf}{
	short = BF,
	long = {brute-force}
}

\DeclareAcronym{peg}{
	short = PEG,
	long = {Parsing Expression Grammar}
}

\DeclareAcronym{dl}{
	short = DL,
	long = {deep learning}
}

\begin{document}

% Title
\title{\doctitle}

% Authors
\author{Marco Edoardo Palma}
\orcid{0000-0003-3300-4828}
\affiliation{%
	\institution{University of Zurich}
	\country{Switzerland}
}
\email{marcoepalma@ifi.uzh.ch}

\author{Pasquale Salza}
\orcid{0000-0002-8687-052X}
\affiliation{%
	\institution{University of Zurich}
	\country{Switzerland}
}
\email{salza@ifi.uzh.ch}

\author{Harald C. Gall}
\orcid{0000-0002-3874-5628}
\affiliation{%
	\institution{University of Zurich}
	\country{Switzerland}
}
\email{gall@ifi.uzh.ch}

\renewcommand{\shortauthors}{Palma et al.}

% Rest of the front
\begin{abstract}
With the presence of online collaborative tools for software developers, source code is shared and consulted frequently, from code viewers to merge requests and code snippets.
Typically, code highlighting quality in such scenarios is sacrificed in favor of system responsiveness.
In these on-the-fly settings, performing a formal grammatical analysis of the source code is not only expensive, but also intractable for the many times the input is an invalid derivation of the language.
Indeed, current popular highlighters heavily rely on a system of regular expressions, typically far from the specification of the language's lexer.
Due to their complexity, regular expressions need to be periodically updated as more feedback is collected from the users and their design unwelcome the detection of more complex language formations.
This paper delivers a deep learning-based approach suitable for on-the-fly grammatical code highlighting of correct and incorrect language derivations, such as code viewers and snippets.
It focuses on alleviating the burden on the developers, who can reuse the language's parsing strategy to produce the desired highlighting specification.
Moreover, this approach is compared to nowadays online syntax highlighting tools and formal methods in terms of accuracy and execution time, across different levels of grammatical coverage, for three mainstream programming languages.
The results obtained show how the proposed approach can consistently achieve near-perfect accuracy in its predictions, thereby outperforming regular expression-based strategies.
\end{abstract}

\begin{CCSXML}
<ccs2012>
<concept>
<concept_id>10010147.10010257.10010293.10010294</concept_id>
<concept_desc>Computing methodologies~Neural networks</concept_desc>
<concept_significance>500</concept_significance>
</concept>
<concept>
<concept_id>10011007.10010940.10010992.10010998.10011000</concept_id>
<concept_desc>Software and its engineering~Automated static analysis</concept_desc>
<concept_significance>300</concept_significance>
</concept>
</ccs2012>
\end{CCSXML}

\ccsdesc[500]{Computing methodologies~Neural networks}
\ccsdesc[300]{Software and its engineering~Automated static analysis}

\keywords{\dockeywords}

% Hacks for author version
\setcopyright{none}
\settopmatter{printacmref=false}
\renewcommand\footnotetextcopyrightpermission[1]{}
\settopmatter{printfolios=true}
\pagestyle{plain}

\maketitle

% Watermark
\watermark{%
This is the authors' version of the paper that has been accepted for publication in the\\%
ACM Joint European Software Engineering Conference and Symposium on the Foundations of Software Engineering (ESEC/FSE 2022)%
}

% Sections
\section{Introduction}
\label{sec:introduction}

% [Context]: services show code to users and what SH is.
Today, software developers often turn to online web applications for support on several aspects concerning their source code manipulation tasks.
Source code repository hosting services, \eg, \textsc{GitLab}, \textsc{BitBucket}, are typically concerned with managing version control instances, DevOps lifecycles, code reviews, continuous integration, and deployment pipelines.
Some extend these functionalities by including issue tracking, knowledge bases, and chats, among other non-software-related features.
Also, some Q\&A platforms, \eg, \textsc{StackOverflow}, provide the possibility to query the community about code-related issues.

With the ability to boost productivity~\cite{sarkar_impact_2015}, code \ac{sh} is popular in online scenarios such as these described.
Formally, \ac{sh} is a form of secondary notation in which portions of the text are displayed in different colors, each representing some feature of the language.
Due to the majority of features only being inferable from the grammatical structure of the input, the task of deciding what color should annotate what portion is non-trivial.
Therefore, resolvers infer the color assignments from some internal grammatical representation of the code.
Intuitively, the more this analysis restricts the belonging of a subsequence to some grammatical productions, the higher is the accuracy of its computation.
As a result of the higher the number of such productions it can recognize, and therefore annotate, the higher is the strategy's coverage.

% Challenge (1)
% [Problem]: challenges in highlighting
Unfortunately, there are two main challenges in performing such analysis in this context.
First, there is a varying level of grammatical validity of the code highlighted.
Due to online code being embedded in multiple contexts, its grammatical correctness cannot be guaranteed.
Indeed, although in version control iterations source code might tend towards being of higher quality, in other cases, such as discussions in code review or chats, this might not carry a valid language derivation, \ie, an \ac{ast} might not be derivable~\cite{terragni_csnippex_2016, jeon_framework_2006, tavakoli_improving_2016, ponzanelli_improving_2014, terragni_apization_2021}.
This inherently induces \ac{sh} strategies in being less reliant on the ability to derive a complete and well-formed representation of the code.

% [Most expensive solution]: Brute-force as most accurate but heaviest.
A \ac{bf} approach towards performing accurate \ac{sh} is to use the language's grammar for the derivation of \acp{ast}, binding a color to each token, based on its location in the tree.
However, not only is this often a computationally expensive strategy, but it also cannot be easily ported to effectively or deterministically recover errors in scenarios of severely incorrect or incomplete language derivations, \eg, code snippets~\cite{medeiros_syntax_2018, demedeiros_automatic_2018, demedeiros_automatic_2020}.
Also, given the rich syntax of modern mainstream programming languages, parsing strategies better suited for dealing with noisy language derivations, \eg, \emph{island parsing}, would expect developers of \ac{sh} tools to produce viable encodings of the languages' original grammars~\cite{moonen_generating_2001, moonen_lightweight_2002}, while still requiring to execute a parsing routine.

\begin{figure*}[tb]
    \centering
    \begin{subfigure}[b]{0.49\linewidth}
        \centering
        \includegraphics[width=\linewidth]{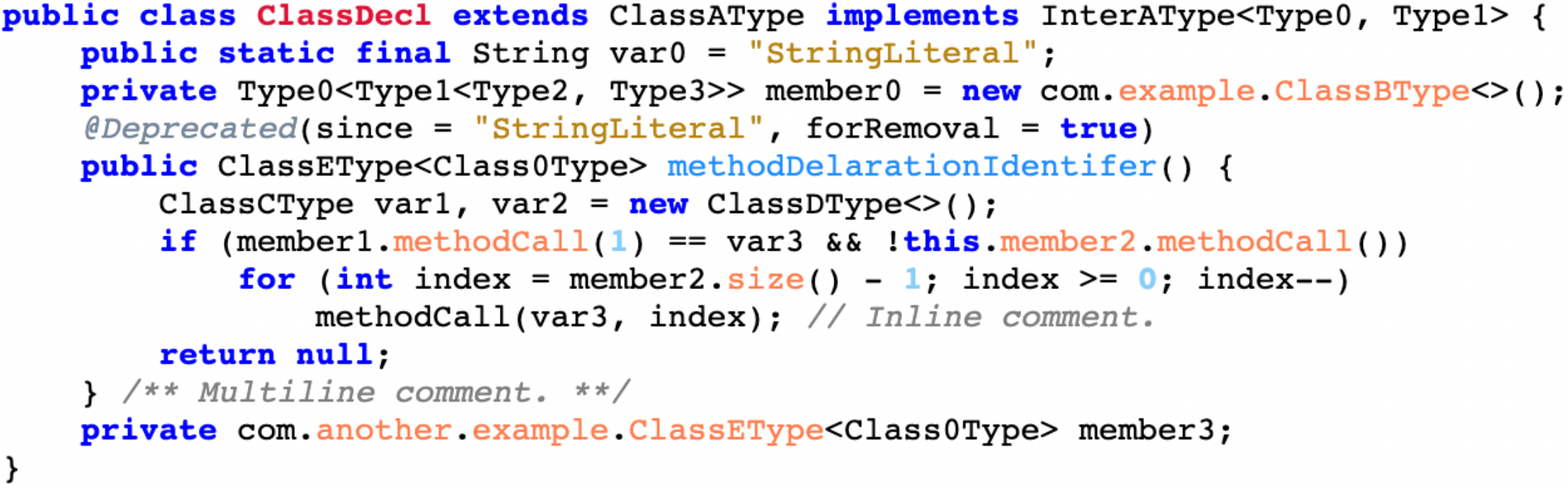}
        \caption{\pygments, \SI{64.76}{\percent} character accuracy.}
        \label{fig:demo:java:pygments}
    \end{subfigure}
    \hfill
    \begin{subfigure}[b]{0.49\linewidth}
        \centering
        \includegraphics[width=\linewidth]{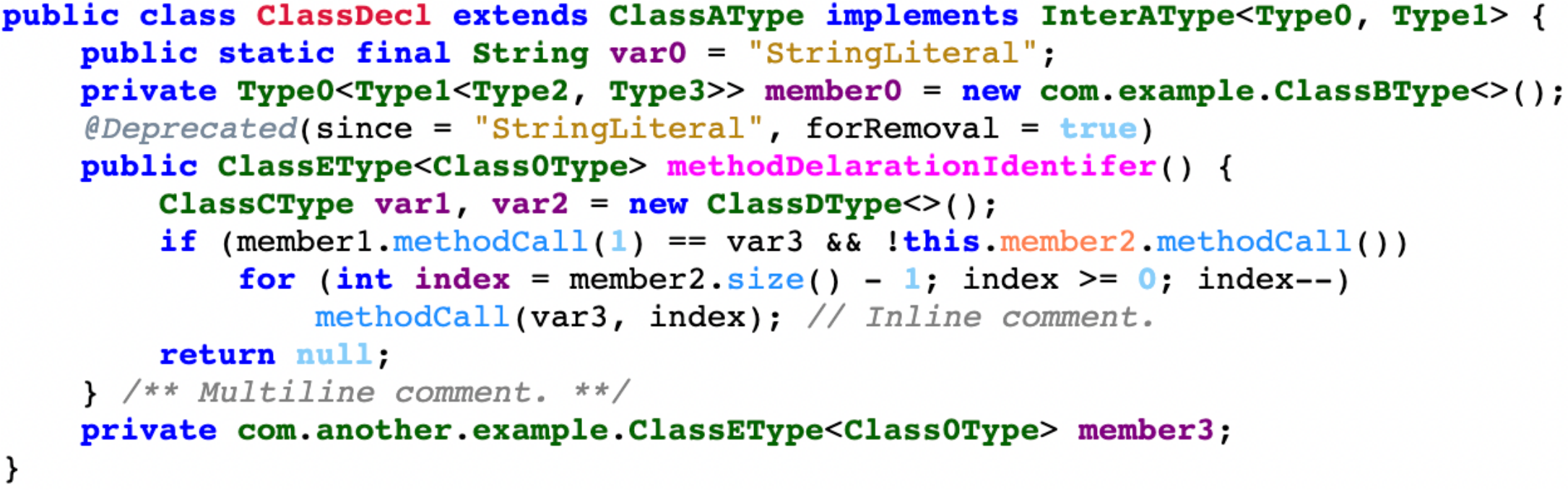}
        \caption{\apprbrnn{16}, \SI{100.00}{\percent} character accuracy.}
        \label{fig:demo:java:brnn}
    \end{subfigure}
    \caption{An example of \begin{revised}task \task{4} of\end{revised} \java \ac{sh}, using the state-of-practice and proposed approaches.}
    \label{fig:demo:java}
\end{figure*}

% Challenge (2)
% [SOTA]: regular expressions fast but inaccurate
As for the second challenge, only a small time delay is allowed for this frequent process to terminate, which \ac{bf} approaches might exceed.
\enquote{On-the-fly} \ac{sh} refers to code being highlighted as this is being retrieved by the user.
The adherence to such computational schema results in \ac{sh} resolvers having an indirect impact on user experience~\cite{hoxmeier_system_2000, kang_experience_2016}.
For the above-mentioned reasons, state of practice \ac{sh} strategies are mainly built around \emph{(per-language) ad-hoc} lexers, which heavily rely on systems of \acp{re}.
Such design allows to achieve excellent computational performances whilst providing a \ac{sh} capable of capturing some contained number of grammatical structures and accuracy levels.
An example of the effects of this strategy is visible in \cref{fig:demo:java}.
Here, the \ac{sh} produced by a popular \acp{re}-based resolver (\cref{fig:demo:java:pygments}) is compared to one producing perfect highlighting for a grammatical coverage resembling those found on \acp{ide} (\cref{fig:demo:java:brnn}).
The low coverage of the former is perceived by its inability to detect identifiers for types, method and variable declarations.
In addition, it cannot distinguish severely distant grammatical constructions such as field accesses, method invocations, and reference types.
In turn, this contributes to low annotation accuracy.
Moreover, the specification of these lexers is often far from that of the language, inducing a tedious and error-prone \acp{re} design process, with the generalizability of the final product relying on the manual compilation of test cases and multiple iterations of user feedback.
%Furthermore, some limitations to the capturing of complex grammatical structures might persist, as the more the \acp{re} are designed to grammatically consume portions of the input, the more backtracking of rules would be required, leading to some, perhaps undesired, increases in computational complexity.

% [Ideal solution]: characteristics
Therefore, motivated by these challenges and shortcomings, it is desirable to have an approach that is:
\begin{inparaenum}[(1)]
    \item \emph{simple to implement}, providing a deterministic, reusable, and low-effort process for developers to create and customize highlighters;
    \item \emph{able to reach high grammatical coverages}, enabling efficient highlighting of more complex grammatical structures than those computed in nowadays online highlighters;
    \item \emph{highly accurate}, closely reproducing the highlighting accuracy of a formal \ac{ast} analysis process;
    \item \emph{input flexible}, reaching high accuracy on correct and incomplete/invalid derivations of the target programming language.
\end{inparaenum}

% [Our solution]
This paper proposes a solution that exploits lightweight \acp{rnn} models to encode the highlighting behaviors of formal \ac{sh} \acf{bf} methods.
A \ac{bf} approach is user-defined and exploits the language's existing lexing and parsing tools to assign each token in the source code to a \ac{sh} class, \ie, an abstraction of the \ac{sh} color, based on its location in the \ac{ast}.
Therefore, it is a formal process, which, if well-formed to match the intended highlighting scheme, is always guaranteed to generate the correct \acp{sh} for files carrying a valid derivation of the language.
After having used \ac{bf} to compile highlighting assignments (\ac{sh}) for multiple sample files, an \ac{rnn} is trained to bind sequences of tokens to sequences of \ac{sh} classes.
\begin{revised}
The training process only occurs once and produces an \ac{rnn} model that is reusable for all future \ac{sh} tasks.
For the training hardware used for the experiments in this work, all the proposed models can be trained in the order of minutes, and comfortably within the one hour mark.
In the case of the non-bidirectional flavors, the delay is cut in half compared to their bidirectional counterpart.
This delay substitutes to today's \emph{state-of-practice} resolvers which involve the development of tedious systems of \apprregex.
\end{revised}
Once trained, the \ac{rnn} computes the \ac{sh} of source code by inferring \ac{sh} classes to the token stream produced by the language's original lexer.

This novel approach to on-the-fly \ac{sh} is tested with regards to its accuracy across four types of grammatical coverages, exploring the detection of different combinations of lexical features and various grammatical constructions for identifiers, declarations, and annotations.
To support the suitability of the proposed approach in the deployment scenarios previously envisioned, this is also tested with regards to its execution time when predicting.
Moreover, the same metrics are measured across three mainstream programming languages: \java, \kotlin, and \python.
All the metrics are also computed for a highly popular \ac{sh} tool, \ie, \pygments~\cite{pygments}, based on the well-establish \ac{re} strategy used by a large number of online vendors such as \textsc{GitLab}, \textsc{BitBucket}, and \textsc{Wikipedia}.
\smallskip

\noindent
To summarize, the main contributions of this paper are:
\begin{itemize}
    \item a dataset for \ac{sh} benchmarking for three popular programming languages, \ie, \java, \kotlin, and \python, obtained through formal \ac{bf} strategies;
    \item the design of an \ac{nn}-based approach for \ac{sh}, with near-perfect highlighting accuracy and suitable prediction delays;
    \item the comparison with the state of practice \ac{sh} strategy in terms of accuracy, coverage, and execution time;
    \item the performance analysis of the approaches in case of incorrect/incomplete source codes.
\end{itemize}
\smallskip

\noindent
The implementation, benchmark datasets, and results are available in the replication package~\cite{replicationpackage} and published at the address \url{\replicationurl}.
\smallskip

\noindent
The rest of this paper is structured as follows.
% In \cref{sec:background}, the background information is introduced.
\Cref{sec:approach}, presents the design of the approach.
\Cref{sec:experiments} describes the experimental setup, whereas \cref{sec:results} shows and discusses the results.
\Cref{sec:related_work} surveys the related work, and \cref{sec:conclusions} concludes with a summary of the findings and contributions, as well as an outlook on future research in this area.

\section{Approach}
\label{sec:approach}

The strategy designed to tackle the challenges raised in this paper aims at deriving \acfp{nn} capable of statistically inferring the perfect behavior of \acf{bf} models.
For this purpose, an oracle of \ac{sh} solutions is generated using the language's \ac{bf} resolver.
The following section presents in detail the specification of both \ac{bf} and \ac{nn} models, as well as providing some motivations for the design.

\subsection{Oracles for Syntax Highlighting}
\label{sec:approach:bf}

\emph{\Acf{bf}} refers to the deterministic process of producing the correct token classification, or \acf{sh}, for some language derivation from which an \acf{ast} is derivable.
These are the sole components for the generation of the \ac{sh} oracle and are created by reusing the language's existing lexing and parsing tools.
The two components respectively represent the input source code as a token stream and order them into an \ac{ast}.
Subsequently, a tree walker exploits the structural information of the \ac{ast} to assign each token to its \ac{sh} class.
This process assumes that a \ac{bf} resolver is guaranteed to compute the correct \ac{sh} of any valid input file, hence setting the highest achievable highlighting accuracy for any coverage specification.
It is important to note that such a design merely requires the developer to implement a walker consisting of only a handful of detection rules, as reported in the replication package~\cite{replicationpackage}.
As a result, the process of producing a \ac{bf} highlighter is deterministic and only asks for a basic understanding of the language's grammar, as it already exists.
It is a significant departure from the tedious and error-prone workflow of defining systems or regular expressions.

The \ac{bf} algorithm is integral in the generation of the \emph{oracle}, \ie, a collection of language's source code files and respective \ac{sh}.
For this purpose, each sample file is piped through the language's lexer and then tokenized.
From each token a new entity is derived in the form: $\textit{\acs{eta}} = \{i_s, i_e, t, tr\}$ where the \emph{\acf{eta}} object is a tuple of:
\begin{inparaenum}
    \item $i_s$ and $i_e$, denoting the token's character start and end indexes respectively, according to the file that contains it;
    \item $t$, the exact text the token references;
    \item \emph{tr}, the token's \emph{Token Rule}, encoded as a natural number, or in other words, the \emph{ID} the language's lexer consistently assigns, through a dictionary, to tokens of the same type, among all types defined in the lexer (\eg, a token of text \code{\{} might corresponds to a lexer type \code{OPEN\_BRACE} hence to the token rule, or unique \emph{ID}, \code{20}).
\end{inparaenum}
For example, \code{String lang = "Java.";} might result in the set of \ac{eta}: \code{\{0, 5, String, 102 \}}, \code{\{7, 10, lang, 102\}}, \code{\{12, 12, =, 73\}}, \code{\{14, 20, "Java.", 55\}}, and \code{\{21, 21, ;, 63\}}.
This representation allows the generalization of \ac{sh} patterns based on the sequence of language features in the form of token types.
It does it by abstracting away the otherwise \enquote{noise}, injected by the tokens' specific text features, transparent to the parsing of the file.
% At this stage, it is important to note that a common practice used in lexing strategies is to skip, or drop, all of those tokens which are not included in any grammar production (\eg redundant whitespace, comments).
% These should be retrieved whenever they are expected to be highlighted: token channels or filtering ahead of parsing are two viable solutions.

Subsequently, the language's parser organizes the tokens into an \ac{ast}.
%Tokens are later ordered into an \ac{ast} according to the language's grammar specification.% and the available error recovery logics.
Walking the \ac{ast} through patterns such as, \emph{Visitor} or \emph{Listener}, all previously computed \acp{eta} are mapped to \emph{\ac{heta}} objects.
These extend \acp{eta} to include a \emph{Highlighting Class} \emph{hc}, corresponding to the grammatical \ac{sh} class to which the token being referenced is part of.
Tokens that are not part of any grammatical construction are bounded to the unique \emph{hc} \code{ANY}, representing text, \ie, no highlighting.
As a result $\textit{\acs{heta}} = \{i_s, i_e, t, tr, hc\}$.
Continuing on the above-mentioned example, the following \ac{heta} set might be computed as:
\code{\{0, 5, String, 102, 1\}}, \code{\{7, 10, lang, 102, 2\}}, \code{\{12, 12, =, 73, 0\}}, \code{\{14, 20, "Java.", 55, 3\}}, \code{\{21, 21, ;, 63, 0\}}, where \emph{hc} of: \code{0} decodes to some not highlighted tokens, \code{1} to type identifiers, \code{2} to variable declaration identifiers, and \code{3} to string literals.

A \ac{bf} resolver for some language $L$ is a function of the form:
${bl}_L: \{c\}, {le}_L, l_L, p_L, {ws}_L \rightarrow \{\textit{\acs{heta}}\}$, where:
\begin{inparaenum}
    \item for the \emph{lexer encoder} ${le}_L: l_L, \{c\} \rightarrow \{\textit{ETA}\}$,
    \begin{inparaenum}[({1}.1)]
        \item $l_L$ is the lexer of $L$,
        \item $\{c\}$ is the character set of the input file,
        \item $\{\textit{\acs{eta}}\}$ the resulting set of \acp{eta}.
    \end{inparaenum}
    \item for the {\it parser} $p_L: l_L, \{c\} \rightarrow {\acs{ast}}_L$, ${\ac{ast}}_L$ is the derived \ac{ast} of the input file,
    \item and for the {\it walking strategy} $ws_L: {\ac{ast}}_L, \{ETA\} \rightarrow \{\textit{\acs{heta}}\}$, $\{\textit{\acs{heta}}\}$ is the oracle for the input's file
\end{inparaenum}

\subsection{\acsp{rnn} for Syntax Highlighting}

In order to efficiently perform \ac{sh} for a given file, this approach seeks to obtain a \acf{nn} model capable of mapping a sequence of token rules $\{tr\}$ to a sequence of \ac{sh} classes $\{hc\}$, as performed by some \ac{bf} resolver.
Hence, the process of computing \ac{sh} becomes a statistical inference on the expected grammatical structure of the token sequence in input.

The motivation behind the use of \acp{nn} for such a task relies on the highly structured nature of programming languages' files.
Indeed, the flow of the incoming characters is:
\begin{inparaenum}
    \item represented as an entity stream selected from a finite set of terminal symbols $\{tr\}$, and
    \item ordered by an underlying pure ordering function as a formal grammar.
\end{inparaenum}
\Ac{sh} can be viewed as the grammar for which there always exists a correct language derivation whenever there exists a valid derivation of the original grammar.
This is true as for some grammar $g$, its highlighter is the grammar $hg$ that sequentially parses sub-productions $s_{hg}$ of $g$, which are enough to discriminate a \emph{tr} subsequence to some target highlighting construction; or otherwise, map every token not consumable by any $s_{hg}$ to a terminal symbol.
In this novel approach to \ac{sh}, the effort of producing such \ac{sh} grammar is lifted from the shoulders of the developers and instead delegated to the \ac{nn} which infers it from the behavior observed from some \ac{bf}.
The task of \ac{sh} is reduced to a \enquote{sequence-to-sequence} translation task~\cite{sutskever_sequence_2014}, \ie, from $\{tr\}$ to $\{hc\}$.

To tackle this new problem reduction, the following proposes the use of \acfp{rnn}~\cite{cho_learning_2014}, for the learning of \ac{sh} sequence bindings.
These offer a base approach to sequence translation by iterating through each value of the input sequence while outputting a unit of translation and carrying forward differentially optimized information to aid the prediction of future inputs.
Furthermore, for those grammars producing sequence distributions for which the binding of an \emph{hc} for some \emph{tr} may require the look ahead of an arbitrary number of tokens, this approach resorts to the use of \acfp{brnn}~\cite{schuster_bidirectional_1997} in place of traditional \acp{rnn}.
Indeed, these also aim at addressing this specific issue by behaving as traditional \acp{rnn}, however inferring the translation of each input from the extra information carried from navigating the input sequence in reverse.
Finally, the model is designed to output for each \emph{tr}, a categorical probability distribution over the set of available \emph{hc}.
The absolute values of such distributions are normalized by a \emph{softmax} function, resulting in the sequence of \emph{hc} for some sequence of \emph{tr} being the set of \emph{max} values of the distribution computed for each \emph{tr}.
Consequently, with regards to \ac{sh}, an \ac{rnn} model $M$ is a function of the form: $M: \{tr\} \rightarrow \{hc\}$.

Although base \acp{rnn} are no longer the state of the art in many translation applications, with current solutions mostly utilizing convolutional layers, the encoder-decoder architectures, or relying on the attention mechanism~\cite{sutskever_sequence_2014, bahdanau_neural_2015, luong_addressing_2015, wu_google_2016, gehring_convolutional_2017, artetxe_unsupervised_2018}, these still offer a lightweight model compared to more recent techniques.
Moreover, as it is later shown, the number of well-formed structural features \ac{nn} are expected to infer from the \ac{sh} oracle samples is small.
It means that the extra infrastructure of deeper networks would result in no appreciable \ac{sh} accuracy increases, but rather in computational overheads and non-trivial hyperparameter/training configurations.
Instead, \acp{rnn} and \acp{brnn} provide a baseline solution for this novel challenge, delivering predictions with contained overheads.
In addition, the training behavior of such models allows this approach to maintain a constant training configuration.
Not only does it result in stable performances across different languages and coverage settings, but also in a solution that is accessible to a broader audience of developers~\cite{anand_black_2020}.

\section{Experiments}
\label{sec:experiments}

The effectiveness of this proposed approach is evaluated in terms of its prediction accuracy and speed for four types of \ac{sh} coverage.
Moreover, in the interest of providing a clearer view on how the performances of this approach might generalize, all experiments were conducted on three mainstream programming languages: \java, \kotlin, and \python.
To represent the state of practice approach using \acp{re}, the \pygments \ac{sh} library~\cite{pygments} is also evaluated against the same metrics.
\pygments is highly popular in online and offline scenarios and found in an array of tools such as \textsc{GitLab}, \textsc{BitBucket}, and \textsc{Wikipedia}.
%This is to contextualize the performance of this novel method with the nowadays' state of practice in this field.
The following research questions are considered for the formal analysis of the solution:
\smallskip

\begin{reqs}
    \item [\req{1}] How accurately can the proposed \ac{nn} approach replicate the \ac{sh} behaviour of a \ac{bf} model?
\end{reqs}
This question aims at evaluating, in terms of \ac{sh} accuracy, for all the defined coverage levels, to what extent the proposed approach can be a substitute to pure \acl{bf} methods.
\smallskip

\begin{reqs}
    \item [\req{2}] How does the proposed \ac{nn} approach compare to nowadays state of practice, or \ac{re}, approaches?
\end{reqs}
This question needs the computation of the \ac{sh} accuracy, for all the defined coverage levels, to understand to what extent the proposed approach can be a substitute to the state of practice.
\smallskip

\begin{reqs}
    \item [\req{3}] How do the speed of computation of the three approaches, \ac{nn}, \ac{bf} , and \ac{re} compare?
\end{reqs}
It provides insights into the time delays required when performing \ac{sh} with the proposed \acp{nn}, \ac{re}-based, and \ac{bf} approaches.
\smallskip

\begin{reqs}
    \item [\req{4}] How accurately can the proposed \ac{nn} approach perform \ac{sh} of incomplete language derivations, compared to the \ac{re} and \ac{bf} approaches?
\end{reqs}
An advantage of both the proposed and \ac{re}-based approaches is their natural portability to estimate \ac{sh} schemes for incorrect/incomplete sequences of tokens.
Hence, this question evaluates, in terms of accuracy, for all the defined coverage levels, how these approaches compare to the theoretical perfect \ac{sh} solution.

\begin{table*}[tb]
    \caption{Metrics for \java, \kotlin, and \python normalized \ac{sh} oracles}
    \label{tab:results:oracle_file_stats}
    \centering
    \resizebox{1.0\linewidth}{!}{
    \rowcolors{2}{gray!10}{}
\begin{tabular}{
    l
    S[table-format=4] S[table-format=5] S[table-format=1] S[table-format=4] S[table-format=7]
    S[table-format=4] S[table-format=5] S[table-format=1] S[table-format=4] S[table-format=7]
    S[table-format=4] S[table-format=5] S[table-format=1] S[table-format=4] S[table-format=7]
}

\hiderowcolors
\toprule

\multirow{2}[2]{*}{\textbf{Metric}} & \multicolumn{5}{c}{\textbf{\java}} & \multicolumn{5}{c}{\textbf{\kotlin}} & \multicolumn{5}{c}{\textbf{\python}} \\
\cmidrule(lr){2-6} \cmidrule(lr){7-11} \cmidrule(lr){12-16}
& {\textbf{Mean}} & {\textbf{SD}} & {\textbf{Min}} & {\textbf{Median}} & {\textbf{Max}} & {\textbf{Mean}} & {\textbf{SD}} & {\textbf{Min}} & {\textbf{Median}} & {\textbf{Max}} & {\textbf{Mean}} & {\textbf{SD}} & {\textbf{Min}} & {\textbf{Median}} & {\textbf{Max}} \\

\midrule
\showrowcolors

Chars & 6239 & 11575 & 0 & 2932 & 504059 & 2455 & 4385 & 80 & 1490 & 176176 & 7390 & 34324 & 0 & 3398 & 3987090 \\
Whitespaces & 1207 & 2417 & 0 & 529 & 72702 & 575 & 1276 & 6 & 282 & 47495 & 1999 & 12941 & 0 & 829 & 1465856 \\
Lines & 190 & 332 & 0 & 94 & 14628 & 70 & 121 & 1 & 43 & 4734 & 208 & 873 & 0 & 104 & 89373 \\
Tokens & 882 & 1745 & 1 & 371 & 45229 & 737 & 1559 & 23 & 327 & 72484 & 1161 & 4997 & 1 & 525 & 448562 \\

\bottomrule

\end{tabular}

    }
\end{table*}

\subsection{Coverage Tasks Definition}

Although an infinite number of coverage schemes could be generated and tested for, the initial iteration of this novel approach to \ac{sh} investigates the highlighting of language features as done in the most common \acp{ide} for the selected languages, such as \textsc{IntelliJ IDEA}, \textsc{PyCharm}, and \textsc{Visual Studio Code}.

Each \emph{Coverage Task} (\emph{T}) is therefore created by combining one or many of the following language feature groups.
Each feature represents a unique \emph{hc} (\emph{Highlighting Class}, see \cref{sec:approach:bf}), or in visual terms, a color.
\smallskip

\noindent
\code{Lexical}:
this group includes token classes that are lexically identifiable, meaning that for a given token, nothing but its \emph{tr} value is required to bind it or not to any of such classes:
\begin{itemize}
    \item \code{KEYWORD}, thereby only referring to strong keywords, as soft keywords may also be used as user-defined identifiers in some allowed language contexts. In this class, also tokens of primitive types, \eg, \code{int}, \code{float}, are included if the language identifies them as such;
    \item \code{LITERAL}, any literal value of the language, \eg, numbers (integers, floating, binary, hexadecimals), boolean values (\code{true}, \code{false}), null constants (\code{null}, \code{None});
    \item \code{CHAR\_STRING\_LITERAL}, any user-defined string or character literals, including those part of string interpolation sequences;
    \item \code{COMMENT}.
\end{itemize}

For this group, all classes are assigned using the same criterion that is applied to all the selected programming languages.
\smallskip

\noindent
\code{Identifier}:
the group includes classes for special types of identifiers:
\begin{itemize}
    \item \code{TYPE\_IDENTIFIER}, matching all the identifier tokens within all the languages' productions representing a type entity;
    \item \code{FUNCTION\_IDENTIFIER}, all the identifiers used in function or methods calls;
    \item \code{FIELD\_IDENTIFIER}, referring to those identifiers that the grammars understand being references to an attribute of an object or entity.
    These are usually preceded by a entity navigation operator, \eg, in \java's \code{Object o = a.b.c().d;}, \code{b} and \code{d} are such \code{FIELD\_IDENTIFIER}, whereas \code{c} might be considered a \code{FUNCTION\_IDENTIFIER}.
\end{itemize}
\smallskip

\noindent
\code{Declarator}:
it includes classes for the classification of token identifiers that carry the name of new top-level features of programs:
\begin{itemize}
    \item \code{CLASS\_DECLARATOR}, referencing identifiers bounded to some newly defined declaration of any form of class, objects, enumerations, data classes, structures, etc.;
    \item \code{FUNCTION\_DECLARATOR}, for identifiers bounded to some newly defined method or function;
    \item \code{VARIABLE\_DECLARATOR}, to some newly defined variable.
    Note the exclusion of this class from \python experiments due to its intrinsic ambiguity of value to identifier assignments.
\end{itemize}
\smallskip

\noindent
\code{Annotation}:
this includes the base annotation components:
\begin{itemize}
    \item \code{ANNOTATION\_DECLARATOR}, as it is common practice to markup annotations in all three selected languages, this class references the token identifiers and prefixed symbols such as the \code{@}, of an annotation.
\end{itemize}
\smallskip

\noindent
Finally, \emph{hc} \code{ANY} gathers all tokens not belonging to any of the categories mentioned above.%; in other words, this is the class of non highlighted text.
\smallskip

\noindent
From the \emph{hc} groups, four coverage tasks are defined to evaluate the flexibility of the \ac{rnn} approach to comply with some arbitrary \ac{sh} coverage.
The four \emph{Coverage Tasks} are defined to demand the identification of the following groups:
\begin{itemize}
    \item \task{1}: $\{$\code{ANY}$\}$, \code{Lexical}, and \code{Declarator};
    \item \task{2}: $\{$\code{ANY}$\}$, \code{Lexical}, and \code{Identifier};
    \item \task{3}: $\{$\code{ANY}$\}$, \code{Lexical}, \code{Declarator}, and \code{Identifier};
    \item \task{4}: $\{$\code{ANY}$\}$, \code{Lexical}, \code{Declarator}, \code{Identifier}, and \code{Annotation}.
\end{itemize}

It is important to note that, for the reported tasks configuration, given the oracle $O_\text{\task{4}}$ carrying all of the language classification groups, the oracle of any other class $O_\text{\task{[1..3]}}$ can be derived directly from $O_\text{\task{4}}$ through means of a \emph{Task Adapter} $TA_\text{\task{4}, \text{\task{[1..3]}}}$.
For any task $O_{\text{\task{i}}\ |\ i \in \{1..3\}}$ a $TA_{\text{\task{4}}, \text{\task{i}}}$ maps every target class \emph{hc} to itself if it is a possible target class for \task{i}, otherwise to the \emph{hc} class \code{ANY} (text).

More details about the above language groups, and their detection strategy for all three investigated languages, are available in the replication package~\cite{replicationpackage}.

\subsection{Data Collection and Preprocessing}

The following describes the procedure produce the datasets used in the experiments.
The full details, together with the downloadable data, are available in the related replication package~\cite{replicationpackage}.

\paragraph{Data mining}
In order to generate \ac{sh} oracles for testing the approaches with regards to their accuracy, speed of evaluation, and training of the \ac{rnn} models, samples for the three programming languages selected are mined from \github's public repositories, through \github's \ac{api}.
In this process, the repositories are pulled by filtering per programming language and sorting by descending order of stars rating.
%This is achieved by setting the query field: \nl{language} to any of \nl{Python}, \nl{Java}, and \nl{Kotlin} strings; hence obtaining the desired ordering by mapping the field \nl{sort} to \nl{stars} and \nl{order} to \nl{desc}.
%Given a repository, files are sampled from the last commit on the \emph{main} branch, and those that match the language's file extension are gathered sequentially, in their natural order.
For every main branch, files matching the language's file extension are downloaded in their natural order.
% Any repository or file that fails to be obtained due to unexpected \ac{api} error is skipped, and the incident is logged.

With the ultimate goal of converting each file to its equivalent set of \ac{heta}, the data collection process filters only files for which the \ac{bf} strategy can derive an \ac{ast}.
Of all files, only one instance of the same token rule (\emph{tr}) sequence is kept: this prevents giving an advantage to the \ac{rnn} approach, which works at a \emph{tr} sequence level instead of at a character level.
Indeed, two program files might carry different text but share the same structure; notice how these two \python code are structurally equal: \code{a = b.c[3].d()} and \code{u\_field = user.files[0].normalised()}.
%Such filtering process is implemented by deriving the files' \emph{tr} sequences, as computed by the language's lexer and keeping only one of each.

For each programming language, the data collection pipeline runs until it has sampled \num{20000} files.
This sample size is in the interest of creating oracles that are both of large statistically meaning, for the average file contents of each language, but could also allow for the execution of extensive accuracy and performance testing.
%, in tractable time spans.
%Such sample size can also be equal across all three languages, further levelling the playing field of all the approaches for each language, also from the data collection front.
Statistics on the number of characters, whitespace, lines of code, and tokens, of the datasets collected for each language are summarized in \cref{tab:results:oracle_file_stats}.

\paragraph{Brute-Force and Oracle Generation}
\label{subsec:experiments:bfandoracle}
To create an oracle for each language, given a set of valid input files, a \ac{bf} method must be created.
As one of the goals of this proposed approach is to reutilize the existing lexing and parsing strategies, the \textsc{ANTLR4}~\cite{antlr} parser generator tool is used, pooling the respective official \textsc{ANTLR4} lexer and parser grammars of each language.
Using \textsc{ANTLR4} proved to be a winning solution to kick-start the creation of all three oracles.
Not only is it a widely popular parser generator, but also used by official language specifications, such as \kotlin, and benefits of an active community developing grammars for most of the mainstream programming languages.
However, it is essential to note that the operability of this approach does not strictly rely on this particular tool, as any preprocessing program could be used if mildly adapted to output the required and largely generic oracle information.

The obtained lexers and parsers, of which version details are available in the replication package~\cite{replicationpackage}, are kept largely unchanged.
The most significant changes interest the lexers, which were instructed to push the skipped tokens, \eg, comments, through the lexers' hidden channel.
Such a (minor) modification enables the approach to obtain tokens for these otherwise dropped entities, which might still require highlighting, as reported in \cref{sec:approach}.
Should this workflow not be available in a language's parsing implementation, or should its introduction cripple the structure of the parser, tokens can be lexed by a dedicated lexer.
%This is not expected to have an impact on the largely linear time complexity of the lexing strategy, as such tokens are usually designed to be easily discarded during one lexing.

In addition to the pipeline for obtaining the \acp{eta} set and \ac{ast} for a given file, a tree walker is created, which aids the conversion of each \ac{eta} into its grammatically highlighted \ac{heta} derivative.
Although multiple walking strategies are available, for the highlighting of the grammatical features considered in this first iteration of this novel \ac{rnn} approach to \ac{sh}, this can most easily be achieved through the \enquote{listener pattern.}
It limits the process to providing highlighting logic for the productions that are expected to contain tokens belonging to any of the target \ac{sh} classes.
All other tokens are instead implicitly mapped to the \code{ANY} class.
As the reporting of the fine details of such implementations would lead to a large and mainly uninteresting listing of tree analysis rules, this can instead be consulted in the replication package~\cite{replicationpackage}.

For each language, the \ac{bf} methods are created for the coverage specified by task \task{4}.
This leads to the generation of an oracle carrying highlighting targets for each \ac{sh} classes present in any given source files.
The \emph{Task Adapter} method described earlier is therefore used to derive the oracles for the other sub coverages of task \task{1}, \task{2} and \task{3}.
This method not only has no effect on the correctness of the derived oracles but it also avoids the definition of a new tree walker and respective time and space expansion for computing further oracles.

\paragraph{Data organization}
As the proposed \ac{rnn} approach involves the training of \acp{nn}, it is important to report on what strategies are put in place to ensure that not only the generalizability of the solution is verified but that there also exists an unbiased setting when its accuracy is compared with the other approaches.

For these reasons, the oracles are randomly shuffled and then split into three folds.
Folds ensure that \SI{33}{\percent} of the oracle's samples are used for testing only, whereas of the remaining \SI{66}{\percent}, \SI{90}{\percent} is reserved for training and \SI{10}{\percent} for validation.
These three sets never intersect, according to the data collection strategy employed.
Moreover, all folds used in the experiments are constant and persisted.
This helps ensure reproducibility and allows each \ac{rnn} model to be compared when trained on equal datasets.

\paragraph{Incomplete files generation}
\label{subsec:experiments:snippets}
Although both the proposed \ac{rnn} approach and the state of practice, based on \acp{re}, are capable of computing \ac{sh} for incorrect program files, their accuracy in these cases cannot be checked exactly, as deterministic oracles for such files are not always derivable.
For this reason, the focus is shifted from mining for incorrect file derivations towards generating invalid language derivations from the set of valid sampled files.

In order to compare the accuracy of the \ac{sh} computed by the proposed and \ac{re}-based strategies, when fed files carrying incomplete (hence invalid) language derivations against the target \ac{sh} computed by a pure process with access to required extra file structure, the files in each test fold are sampled line-wise to generate one code snippet sized files.
These are drawn from the test datasets, as in this first iteration of this approach, the network is not trained on these incomplete files but only tested; however, sampling for training datasets of the folds might have given an unfair advantage to the \ac{rnn} approach.

At this stage, it is also important to note that it would not have been tractable to sample such snippet-sized file from distributions of natural snippets generating processes, as the \ac{bf} method would not have been available for the formal computation of the correct \ac{sh}, for the reasons highlighted above.
Therefore, with the target number of newly generated files of \num{5000} from each fold test set, thus \num{15000} per language, each test file is drawn randomly, and from it, a random sub-sequence of lines is chosen.

The lengths of the snippets are drawn normally according to the language's mean, standard deviation, minimum and maximum number of snippets lines, determined by number of lines found by querying the \textsc{StackExchange Data Explorer}~\cite{stackexchange}, focusing on snippets from \textsc{StackOverflow}.
In particular, at the time of the experiments, these numbers were (mean, standard deviation, minimum, maximum): \num{17.00}, \num{28.75}, \num{1}, and \num{1117} for \java; \num{15.00}, \num{22.05}, \num{1}, and \num{703} for \kotlin; \num{14.00}, \num{20.39}, \num{1}, and \num{1341} for \python.

Both test files and lines are sampled with replacement.
Given the lines selected, the process gathers the set of \acp{heta} in range and produces a new oracle instance.

\subsection{Compared Approaches}
\label{subsec:experiments:approaches}
Multiple variations of baseline \acp{rnn} models are investigated.

\begin{revised}
An initial configuration for the \acp{rnn} and training was derived by improving the convergence of the networks on the validation set of only the first fold of the \java dataset.
The initial embedding layer was kept at \num{128}, \ie, the smallest power of two larger than the number of token ids for the languages, while the hidden units were added in increasing power of two.
With a constant learning rate of $10^{-3}$ and Adam optimizer, \num{16} and \num{32} (B)\acp{rnn} were found to produce near-perfect accuracy, with the latter not improving in wider models.
Accuracy converged after the second epoch.
A final investigation involved the common practice of reducing the learning rate after convergence by a factor of \num{10}, \ie, $10^{-4}$.
It further helped improve the accuracy of the model, which again was observed to converge within the following two epochs.
\end{revised}

\begin{revised}As a result, the \ac{rnn} models evaluated consist of a fixed \num{128} embedding layer.\end{revised}
The output of the embedding layer is mapped to a single layer \acp{rnn} or \acp{brnn}, of widths evaluated among \num{16} and \num{32} hidden units.
The output of all the \acp{rnn} or \acp{brnn} is passed through a fully connected linear layer reducing it to a categorical distribution of the available \emph{hc}, depending on the \emph{Coverage Task}.
This results in the testing of four models, identified by its directionality and width of \ac{rnn} layer: \apprrnn{16}, \apprrnn{32}, \apprbrnn{16}, and \apprbrnn{32}.
Every model is trained sequentially on each training sample, with cross-entropy loss and Adam optimizer.
The training session for any \ac{sh} \ac{rnn}, language and coverage, was\begin{revised} accordingly\end{revised} set to train for two epochs with a learning rate of $10^{-3}$, and for a subsequent two epochs with a learning rate of $10^{-4}$.
It is in respect of the approach's initial guarantee of delivering a training configuration capable of achieving the performance advertised without the tweaking of the training session by expert developers.
All models commence the training process from a randomly initialized state, according to the deep learning framework utilized, \ie, \textsc{PyTorch}~\cite{paszke_pytorch_2019}, while a constant seed ensures the reproducibility of the experiments.

To contextualize the performances produced by the \acp{rnn} approaches, the vastly popular and well-established \ac{re}-based syntax highlighter \pygments is tested~\cite{pygments} \begin{revised}using its latest available version at the time of testing \emph{2.10.0}\end{revised}.
In the following, \pygments is being referred to as \apprregex.
Its output was manually adjusted to output the same classes included in \task{4}, of which details can be found in the replication package~\cite{replicationpackage}.
Hence, the same \emph{Task Adapter} used during the conversion of the oracle to any other task is used to map each \pygments' prediction to its task-specific class.
%It is important to note that the \apprbf and \apprregex strategies might produce different tokenization of the same file.

Finally, the same \apprbf methods used for the generation of the oracles are reused for the outlined comparisons with the \acp{rnn} and \ac{re}-based approaches.
The use of \textsc{ANTLR4} is not only induced by the large availability of language grammar, but also by its highly efficient \emph{LL(*)} parsing~\cite{parr_adaptive_2014} strategy, and native error recovery logic, both of which undermine the real-world performance advantage of, otherwise theoretically regarded as most efficient, \ac{peg} parsers~\cite{ford_parsing_2004} in both fronts~\cite{medeiros_syntax_2018, hutchison_pika_2020, becket_dcgs_2008}.

\subsection{Evaluation Metrics}
\label{subsec:experiments:metrics}

The quality of an \ac{sh} can be measured with regards to its coverage, accuracy, and speed, described in the following.

\paragraph{Coverage}
The absolute number of unique grammar constructions the highlighter is able to recognize.

\paragraph{Accuracy}
Given a coverage specification, the degree to which the highlighter can bind each character in the input text to its correct \ac{sh} class.
It also resolves the issue of \apprbf and \apprregex strategies possibly producing different tokenizations of the same file.

\paragraph{Speed}
The time delay for the computing of \ac{sh}.
Prediction speed for all methods evaluated during the experimentations is measured as the absolute time in nanoseconds required to predict the \ac{sh} of an input file once this has been supplied.
For each \ac{sh} method, the following time delays are measured:
\begin{itemize}
    \item \apprbf: the time to natively parse the input file and perform a \ac{sh} walk of the obtained \ac{ast};
    \item \apprregex: the time to compute the output vector of \ac{sh} classes, once given the file's source text, but excluding the time required to format the output to any specification.
    The latter is achieved by defining a new \pygments \emph{Formatter} object which accepts the computed \ac{sh}, but does not invest computational time into outputting it, hence removing the added time complexity any specific format might introduce, thereby highlighting the complexity of the approach's underlying \ac{sh} strategy;
    \item \acp{rnn}: the time for the \textsc{ANTLR4} inherited lexer (the same used by the \apprbf approach) to tokenize the input file into a sequence of token rules, plus the time for the \ac{rnn} model to create the input tensor, and predict the complete output vector of \ac{sh} classes.
\end{itemize}

\subsection{Execution Setup}

All \ac{rnn} models are trained on a machine equipped with an AMD EPYC 7702 \num{64}-Core CPU clocked at \SI{2.00}{\giga\hertz}, \SI{64}{\giga\byte} of RAM, and a single Nvidia Tesla T4 GPU with \SI{16}{\giga\byte} of memory.
Instead, all performance testing for all of the compared approaches was carried out on the same machine with an \num{8}-Core Intel Broadwell CPU clocked at \SI{2.00}{\giga\hertz} with \SI{62}{\giga\byte} of RAM.

\subsection{Threats to Validity}
\label{subsec:experiments:threats}
With regards to the problem statement raised in this paper, \ie, on-the-fly \ac{sh}, \textsc{ANTLR4} undoubtedly represents the package of technologies and strategies required not only for the definition of \ac{bf} models but also their evaluation.
% -- as discussed in \cref{subsec:experiments:approaches} and \cref{subsec:experiments:bfandoracle}.
Despite the best intention to consider all viable options, one should not exclude the existence of, perhaps language-specific, parsing tools that might scale the performance of \ac{bf} resolvers.
%However, the real-world requirement for online \ac{sh} imposed on such tools (\cref{sec:introduction}) should be one of the filtering criteria when considering this threat to validity.

The impossibility to generate testing oracles from snippets produced by online user processes, resulted in a first experiment setup which synthetically generates incomplete/incorrect language derivations from the set of parsable derivations.
Therefore, it is crucial to note that \req{4} only intends to provide an initial perspective on how the three approaches might perform on file segments, and at that the formal measure of closeness between this synthetic process to that observable in online code snippets is unknown.
\begin{revised}
Moreover, human annotator processes are likely to employ their statistical inference about the missing context of some code fragment.
Hence, one may argue that conducting such an assessment with a manually composed, and therefore inconsistent, dataset would instead validate a model's ability to meet the level of \emph{program-comprehension} of the sample of users that created the dataset.
Instead, the synthetic dataset created here indirectly validates the model's ability to infer the statistically most likely missing context.
\end{revised}
%The replication package is however provided for future analysis.

% ? Pygments is the prominent resolver but there are other highlighting libraries for online scenarios

\begin{revised}
\pygments provides syntax highlighting for \num{534} languages.
However, it is a collection of implementations of language-specific \apprregex \ac{sh}, and not a single generic \ac{sh} resolver.
This work compares with three of such highlighters, \ie, \java, \kotlin, and \python, but promises to be applicable to other languages, as language-specific \ac{bf} can be used to train new language models.
The validation of the proposed approach across all the languages supported by the \apprregex-based counterpart would extensively assess the generalizability of the strategy.
Therefore, this aspect is considered a limitation of the experimental setup, which does not prove the absolute generic performances of this novel strategy but instead delivers seminal evidence of its applicability.
\end{revised}

Benchmarks for prediction delays might only give a general perspective of the performances of such tools, but exclude specific implementation optimizations that developers might design.
It may also include file size limits for online consumption, which might be platform dependent.
Other variables might concern the efficiency of the integration of \ac{sh} resolvers with the rest of the service, caching strategies, or hardware specifications.
For example, the proposed \ac{rnn} solution might perform differently if run on more production-focused deep learning libraries~\cite{abadi_tensorflow_2015}, or on GPUs.

% delays might need to be adjusted according to the platform's file size limits (eg. large max values esp for python)
% computation delays included are in the best interest of excluding better implementation/architecture of the various pipelines. this considers a base scenario
% don't exclude different hardware might favour one approach or the other
% proposed approach might perform better if run on gpu and on production focused dl distributions such as tf
% large ram in prediction task to avoid accomodate large memory usage of bf models, in production these might not be

% \paragraph{Internal validity}
% Description.

% \paragraph{External validity}
% Description.

% \paragraph{Construct validity}
% Description.

\section{Results}
\label{sec:results}

Developing from the experiment setups described in \cref{sec:experiments}, this section individually addresses the performance of the proposed approach with regards to the four research questions identified.
For each question, its specific validation workflow is described, and the results are presented and discussed.

To compare the observations, the \enquote{Kruskal-Wallis H} test~\cite{montgomery_design_2017} was applied with the \enquote{Vargha-Delaney $\hat{A}_{12}$} test~\cite{vargha_critique_2000}, for the effect size to characterize the magnitude of such differences.
For this reason, the following reports the evaluation metrics in terms of median values, being these tests based on the median differences.

\subsection{\req{1} -- Comparison with \apprbf's Accuracy}
\label{subsec:results:rq1}

\req{1} aims at evaluating the \ac{sh} accuracy of the proposed approach when compared to the theoretical perfect \ac{bf} resolver, on language derivation for which an \ac{ast} is derivable.
Such aspect is validated regarding all three programming languages, as well as to the four \emph{Coverage Tasks}.
Every candidate \ac{rnn} model is first individually trained on the training set of each fold, and its accuracy is recorded about its predictions on the corresponding test set.

\begin{table*}[tb]
    \caption{Median values over \num{3} folds for the accuracy. The maximum scores per task are highlighted}
    \label{tab:results:rq1}
    \centering
    % \resizebox{0.8\linewidth}{!}{
    \sisetup{table-format=1.4}
\rowcolors{2}{gray!10}{}
\begin{tabular}{
    l SSSS SSSS SSSS
}

\hiderowcolors
\toprule

\multirow{2}[2]{*}{\textbf{Model}} & \multicolumn{4}{c}{\textbf{\java}} & \multicolumn{4}{c}{\textbf{\kotlin}} & \multicolumn{4}{c}{\textbf{\python}} \\
\cmidrule(lr){2-5} \cmidrule(lr){6-9} \cmidrule(lr){10-13}
& {\textbf{\task{1}}} & {\textbf{\task{2}}} & {\textbf{\task{3}}} & {\textbf{\task{4}}} & {\textbf{\task{1}}} & {\textbf{\task{2}}} & {\textbf{\task{3}}} & {\textbf{\task{4}}} & {\textbf{\task{1}}} & {\textbf{\task{2}}} & {\textbf{\task{3}}} & {\textbf{\task{4}}} \\

\midrule
\showrowcolors

\apprregex{} & 0.8662 & 0.7606 & 0.7233 & 0.7230 & 0.8009 & 0.6998 & 0.6787 & 0.6781 & 0.9364 & 0.8189 & 0.8189 & 0.8165 \\
\apprrnn{16} & 0.9987 & 0.9716 & 0.9676 & 0.9668 & \tabhvalue 1.0000 & 0.9627 & 0.9598 & 0.9605 & \tabhvalue 1.0000 & 0.9560 & 0.9559 & 0.9550 \\
\apprrnn{32} & \tabhvalue 1.0000 & 0.9751 & 0.9710 & 0.9706 & \tabhvalue 1.0000 & 0.9648 & 0.9640 & 0.9631 & \tabhvalue 1.0000 & 0.9572 & 0.9571 & 0.9570 \\
\apprbrnn{16} & \tabhvalue 1.0000 & \tabhvalue 1.0000 & \tabhvalue 1.0000 & \tabhvalue 1.0000 & \tabhvalue 1.0000 & \tabhvalue 1.0000 & \tabhvalue 1.0000 & \tabhvalue 1.0000 & \tabhvalue 1.0000 & \tabhvalue 1.0000 & \tabhvalue 1.0000 & \tabhvalue 1.0000 \\
\apprbrnn{32} & \tabhvalue 1.0000 & \tabhvalue 1.0000 & \tabhvalue 1.0000 & \tabhvalue 1.0000 & \tabhvalue 1.0000 & \tabhvalue 1.0000 & \tabhvalue 1.0000 & \tabhvalue 1.0000 & \tabhvalue 1.0000 & \tabhvalue 1.0000 & \tabhvalue 1.0000 & \tabhvalue 1.0000 \\

\bottomrule

\end{tabular}

    % }
\end{table*}

As reported in \cref{tab:results:rq1}, for all the languages and coverage tasks selected in this experiment, the proposed approach is capable of producing near-perfect \ac{sh} solutions.
The bidirectional variants prove to be the most eclectic model, which, even in the narrowest tested configuration (\apprrnn{16}), achieve a perfect score more consistently than any base \ac{rnn} model, across all languages and tasks.
It is as expected, with bidirectionally extending the context around each token.
Hence, it enables the resolution of ambiguous syntactical structures of which type is dependent on the next tokens.
%The \ac{rnn}s instead rely their predictions on the expectations built from the token sequences preceding such depths.

%An intuitive example of such patterns is function identifiers, which in all the languages tested, in some instances can discriminated from other classes, in which an identifier is followed by open parenthesis, and no prior function or constructor declarator parts.
%However, this construct remains undecidable until such following parenthesis is seen.
%Hence a disadvantage for the \ac{rnn} strategy, which is set to classify each input token sequentially.

\begin{figure}[tb]
    \centering
    \includegraphics[width=1.0\linewidth]{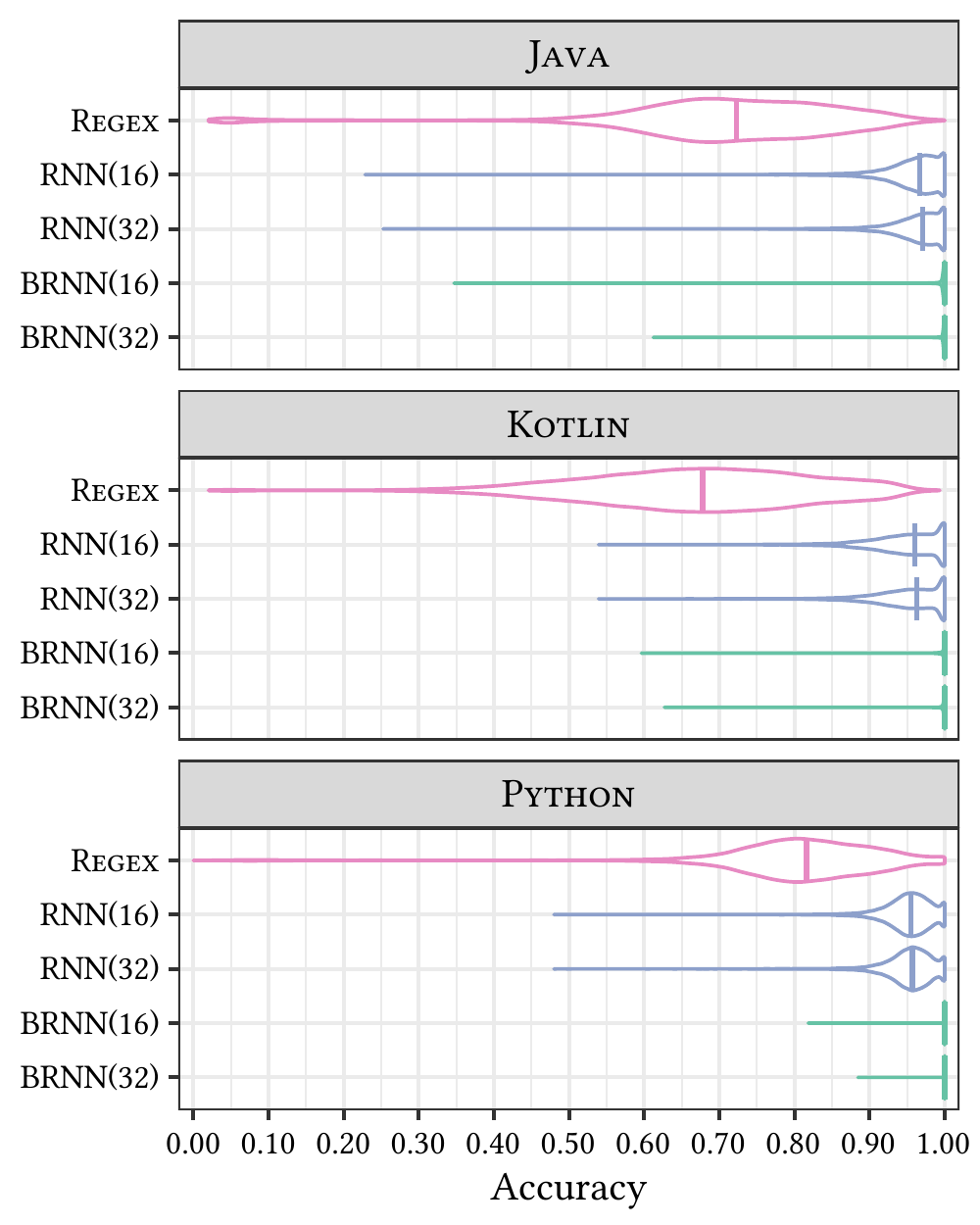}
    \caption{Accuracy values comparison for \task{4}.}
    \label{fig:results:boxplot_rq1}
\end{figure}

Furthermore, the \ac{brnn} variant promotes a significant improvement in the stability of this strategy, with the accuracy distribution more concentrated around the perfect mark and the outliers being not only fewer but also of generally higher accuracy than otherwise obtainable with base \acp{rnn}.
This is clearly visible in \cref{fig:results:boxplot_rq1}.

Therefore, for the average case, the proposed \ac{rnn} strategy to \ac{sh} is most often able to perform as well as the pure \ac{bf} strategy.
Nevertheless, being this a nondeterministic approach, some contained levels of inconsistency should be expected.

\begin{table*}[tb]
    \caption{Descriptive statistics of execution time (ms)}
    \label{tab:results:rq3}
    \centering
    \resizebox{1.0\linewidth}{!}{
    \rowcolors{2}{gray!10}{}
\begin{tabular}{
    l
    S[table-format=3.3] S[table-format=3.3] S[table-format=1.3] S[table-format=2.3] S[table-format=5.3]
    S[table-format=3.3] S[table-format=3.3] S[table-format=1.3] S[table-format=2.3] S[table-format=5.3]
    S[table-format=3.3] S[table-format=3.3] S[table-format=1.3] S[table-format=2.3] S[table-format=5.3]
}

\hiderowcolors
\toprule

\multirow{2}[2]{*}{\textbf{Model}} & \multicolumn{5}{c}{\textbf{\java}} & \multicolumn{5}{c}{\textbf{\kotlin}} & \multicolumn{5}{c}{\textbf{\python}} \\
\cmidrule(lr){2-6} \cmidrule(lr){7-11} \cmidrule(lr){12-16}
& {\textbf{Mean}} & {\textbf{SD}} & {\textbf{Min}} & {\textbf{Median}} & {\textbf{Max}} & {\textbf{Mean}} & {\textbf{SD}} & {\textbf{Min}} & {\textbf{Median}} & {\textbf{Max}} & {\textbf{Mean}} & {\textbf{SD}} & {\textbf{Min}} & {\textbf{Median}} & {\textbf{Max}} \\

\midrule
\showrowcolors

\apprbf{} & 225.684 & 894.046 & 0.004 & 45.903 & 49618.222 & 30.950 & 87.893 & 0.011 & 8.080 & 14119.526 & 52.798 & 242.363 & 0.033 & 24.022 & 23628.056 \\
\apprregex{} & 0.015 & 0.040 & 0.004 & 0.011 & 22.975 & 0.010 & 0.047 & 0.004 & 0.009 & 27.468 & 0.016 & 0.030 & 0.003 & 0.013 & 7.048 \\
\apprrnn{16} & 9.195 & 18.704 & 0.206 & 3.877 & 689.178 & 8.383 & 31.805 & 0.370 & 3.612 & 12755.019 & 66.313 & 288.904 & 0.182 & 32.597 & 27164.357 \\
\apprrnn{32} & 9.202 & 18.581 & 0.195 & 3.887 & 677.833 & 8.439 & 30.231 & 0.384 & 3.666 & 12067.893 & 63.522 & 276.867 & 0.176 & 31.682 & 26279.598 \\
\apprbrnn{16} & 17.506 & 36.176 & 0.270 & 7.241 & 1269.607 & 14.997 & 40.814 & 0.586 & 6.537 & 12120.509 & 75.235 & 333.959 & 0.217 & 35.742 & 32334.076 \\
\apprbrnn{32} & 17.728 & 36.565 & 0.278 & 7.396 & 1341.984 & 15.664 & 42.090 & 0.605 & 6.829 & 12243.090 & 76.895 & 344.068 & 0.219 & 36.301 & 32475.535 \\

\bottomrule

\end{tabular}

    }
\end{table*}

\subsection{\req{2} -- Comparison with \apprregex's Accuracy}
\label{subsec:results:rq2}

Addressing \req{2} allows for the contextualization of the accuracy values obtainable by the proposed strategy, with what is achievable with today's state of practice, \ie, \acp{re}.
%For this purpose, the \pygments library, a widely popular \ac{re}-based tool for \ac{sh} utilised in a great number of online and offline applications, including \emph{GitLab}, \emph{BitBucket}, \emph{Wikipedia}, and many others.
%Furthermore, it has support for a large collection of languages, and its highlighting is language specific; something that is not true for other options which are generic lexers.
Such a research question is therefore tackled by evaluating the \ac{sh} accuracy of \pygments on the same test datasets used to estimate the generalizing accuracy of the \ac{rnn} models in \req{1}.

As supported by the evidence displayed in \cref{tab:results:rq1}, which reports the median accuracy values per \ac{sh} method, the \ac{re}-based strategy consistently performs the worst across all tested scenarios.
%It performs the worst across all coverages in \kotlin, whilst \java sees a noticeable boost followed by \python for which it performs the best.
It is also essential to notice how the \apprregex approach is significantly more prone to variability in its level of accuracy, compared to any of the \ac{rnn} models tested, as visualized in \cref{fig:results:boxplot_rq1}.
\pygments yields its best performance across all languages when its output is evaluated about coverage task \task{1}.

Another observation concerns \pygments's accuracy decaying significantly for all tasks other than \task{1}.
Compared to the other tasks, \task{1} requires the identification of only lexical features and declarator identifiers.
However, unlike declarations, lexical components are always deterministically identifiable through lexing, except soft keywords.
\task{1} is, therefore, the least complex task out of all of those tested as, per file, only a handful of declaration identifiers are found, requiring the resolvers to identify mainly lexical features.
Hence, the accuracy of \apprregex resolver converges considerably for tasks \task{2}, \task{3} and \task{4}, as all other grammatical features are reasonably consistently bounded to incorrect \emph{hc} values.

Overall, the evidence collected for \req{2} supports the fact that the proposed approach is capable of quite consistently boosting the \ac{sh} accuracy otherwise achievable with the state of practice.

\subsection{\req{3} -- Speed Comparison}
\label{subsec:results:rq3}

\begin{figure*}[tb]
    \centering
    \includegraphics[width=1.0\linewidth]{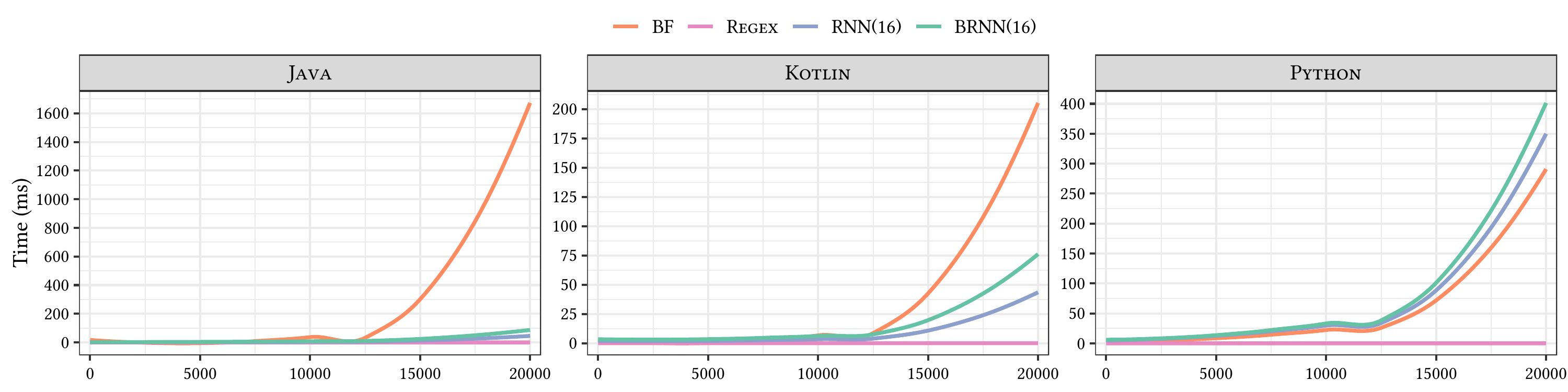}
    \caption{Execution time (ms) values trends comparison for \task{4}.}
    \label{fig:results:boxplot_rq3}
\end{figure*}

The investigation into the prediction speed of all the available approaches aids in contextualizing at what responsiveness costs the proposed approach to \ac{sh} can deliver its coverage and accuracy performances.
Thus, each resolver is set to produce \ac{sh} for each language's oracle \num{30} times, and their prediction delays are recorded.
The experiments are carried out on the same machine, and no GPU is used for the evaluation of the execution of \ac{nn} based resolvers.
From the results obtained and summarised in \cref{tab:results:rq3}, several observations can be made.

The \ac{rnn} based approaches provide significant speed-ups over the \ac{bf} resolvers.
In fact, in the case of \java prediction delays are \num{25} times smaller for \acp{rnn} of both \num{16} and \num{32} hidden units; and \num{13} times smaller in the case of the bidirectional variants.
Moreover, the standard deviation of the prediction delays of the proposed solution is also significantly smaller than the \ac{bf} counterparts.
Both \ac{rnn} models reduce this metric by a factor of \num{38} and the \ac{brnn} models by a factory of \num{25}.
\kotlin leads to similar conclusions, although with the \ac{bf} solution yielding better performances, but still worst compared to the proposed solution.
In particular, the gains in favor of the \ac{rnn} models, which do remain consistent with the delays recorded in \java, decrease to an average speed-up of \num{4} for the \ac{rnn} models, and \num{2} for the \ac{brnn} models.
Standard deviation is also down by a factor of \num{3} and \num{2} for the \ac{rnn} and \ac{brnn} respectively.

Nevertheless, such a narrative changes when comparing the performance of the \ac{nn} approach to the \ac{bf} resolver for \python.
\begin{revised}
According to \cref{tab:results:rq3}, the proposed \ac{rnn} approach is not superior to the \apprbf approach.
In fact, the parsing proves to be significantly more efficient than it is in the cases of \java and \kotlin.
\end{revised}
With the technologies constant for all \ac{bf} resolvers, this suggests the grammar of the \python language is the main promoter for the efficiency gains observed.
Nonetheless, the proposed approach proves capable of nearing such stellar performance of the \ac{bf} resolver, however with some contained slowdowns: \num{1.3} and \num{1.4} on average for the \ac{rnn} and \ac{brnn} models respectively.
Standard deviation is also mildly down by \num{1.2} and \num{1.4} for the \ac{rnn} and \ac{brnn} models.

As expected, the computational overheads of the proposed \ac{nn} approach are more significant compared to the ones that accompany \apprregex.
However, with the \ac{rnn} strategy focused on delivering greater \ac{sh} accuracy and coverage, and a significantly smaller development effort for developers, the focus is shifted on the suitability of this approach to the task.
Considering the average delays recorded during this experiment, these are found to be relatively small.
For the \ac{rnn} approaches predictions are on average delivered in \num{9}\emph{ms}, \num{8}\emph{ms}, and \num{66}\emph{ms}, for \java, \kotlin and \python respectively; and the medians \num{4}\emph{ms}, \num{4}\emph{ms} and \num{33}\emph{ms}.
Such computational delays would most comfortably belong with the \emph{Seow}’s response-time categorization of \emph{instantaneous}~\cite{seow_designing_2008}.
In this category includes human and computer interactions that are expected to complete within \SI{100}{\milli\second} and \SI{200}{\milli\second}, \eg, clicking and typing; whilst longer delays, within \SI{500}{\milli\second} and \SI{1000}{\milli\second}, being categorized as immediate, this last one including navigation actions~\cite{dabrowski_40_2011, seow_designing_2008}.

\Cref{fig:results:boxplot_rq3} shows a smoothed line plot to represent the execution times for all the experiments.
As it shows, the proposed approach is capable of delivering \ac{sh} results well within the average human deadlines, with these requiring delays to be within \SIrange{2}{5}{\second} to maintain flow~\cite{dabrowski_40_2011, seow_designing_2008}, and tolerating a webpage response of \SI{2}{\second}~\cite{nah_study_2004}.

\subsection{\req{4} -- Incomplete Derivations Highlighting}
\label{subsec:results:rq4}

\req{4} considers \ac{sh} accuracy of the highlighters with incomplete/incorrect language derivations.
Likewise, for \req{1} and \req{2}, all approaches are set to produce highlighting for all three languages and four coverage tasks.
The dataset used for this \req{4} is the generated snippet dataset, for which perfect target solutions are known.

\begin{table*}[tb]
    \caption{Median values over \num{3} folds for the accuracy for snippets. The maximum scores per task are highlighted}
    \label{tab:results:rq4}
    \centering
    % \resizebox{0.8\linewidth}{!}{

\sisetup{table-format=1.4}
\rowcolors{2}{gray!10}{}
\begin{tabular}{
    l SSSS SSSS SSSS
}

\hiderowcolors
\toprule

\multirow{2}[2]{*}{\textbf{Model}} & \multicolumn{4}{c}{\textbf{\java}} & \multicolumn{4}{c}{\textbf{\kotlin}} & \multicolumn{4}{c}{\textbf{\python}} \\
\cmidrule(lr){2-5} \cmidrule(lr){6-9} \cmidrule(lr){10-13}
& {\textbf{\task{1}}} & {\textbf{\task{2}}} & {\textbf{\task{3}}} & {\textbf{\task{4}}} & {\textbf{\task{1}}} & {\textbf{\task{2}}} & {\textbf{\task{3}}} & {\textbf{\task{4}}} & {\textbf{\task{1}}} & {\textbf{\task{2}}} & {\textbf{\task{3}}} & {\textbf{\task{4}}} \\

\midrule
\showrowcolors

\apprbf{} & 0.9211 & 0.7421 & 0.6586 & 0.6440 & 0.0000 & 0.0000 & 0.0000 & 0.0000 & \tabhvalue 1.0000 & \tabhvalue 1.0000 & \tabhvalue 1.0000 & \tabhvalue 1.0000 \\
\apprregex{} & 0.8700 & 0.6859 & 0.6346 & 0.6340 & 0.8117 & 0.6577 & 0.6285 & 0.6279 & 0.9338 & 0.7890 & 0.7890 & 0.7860 \\
\apprrnn{16} & \tabhvalue 1.0000 & 0.9582 & 0.9512 & 0.9506 & \tabhvalue 1.0000 & 0.9503 & 0.9469 & 0.9467 & \tabhvalue 1.0000 & 0.9605 & 0.9595 & 0.9587 \\
\apprrnn{32} & \tabhvalue 1.0000 & 0.9634 & 0.9557 & 0.9555 & \tabhvalue 1.0000 & 0.9534 & 0.9513 & 0.9512 & \tabhvalue 1.0000 & 0.9618 & 0.9614 & 0.9617 \\
\apprbrnn{16} & \tabhvalue 1.0000 & \tabhvalue 1.0000 & \tabhvalue 1.0000 & \tabhvalue 1.0000 & \tabhvalue 1.0000 & \tabhvalue 1.0000 & \tabhvalue 1.0000 & \tabhvalue 1.0000 & \tabhvalue 1.0000 & \tabhvalue 1.0000 & \tabhvalue 1.0000 & \tabhvalue 1.0000 \\
\apprbrnn{32} & \tabhvalue 1.0000 & \tabhvalue 1.0000 & \tabhvalue 1.0000 & \tabhvalue 1.0000 & \tabhvalue 1.0000 & \tabhvalue 1.0000 & \tabhvalue 1.0000 & \tabhvalue 1.0000 & \tabhvalue 1.0000 & \tabhvalue 1.0000 & \tabhvalue 1.0000 & \tabhvalue 1.0000 \\

\bottomrule

\end{tabular}

    % }
\end{table*}

As it possible to notice by comparing \cref{tab:results:rq4}, related to \req{4}, with \cref{tab:results:rq1}, related to \req{1}, the results show how the \ac{rnn}-based approaches are capable of maintaining accuracy performances on par with those obtainable on language derivations for which an \ac{ast} is derivable.
In fact, also in this scenario the \ac{rnn} models compute \ac{sh} with an accuracy within \SIrange{94}{96}{\percent}, and the bidirectional variants always reaching a perfect median accuracy value.
The state of practice, \ie, \apprregex, registers a decrease in accuracy, which, similarly to \req{2}, is considerably far from those obtainable with the proposed \ac{nn} models.
It is especially noticeable for tasks with larger grammatical coverage, such as \task{4}.

\begin{figure}[tb]
    \centering
    \includegraphics[width=1.0\linewidth]{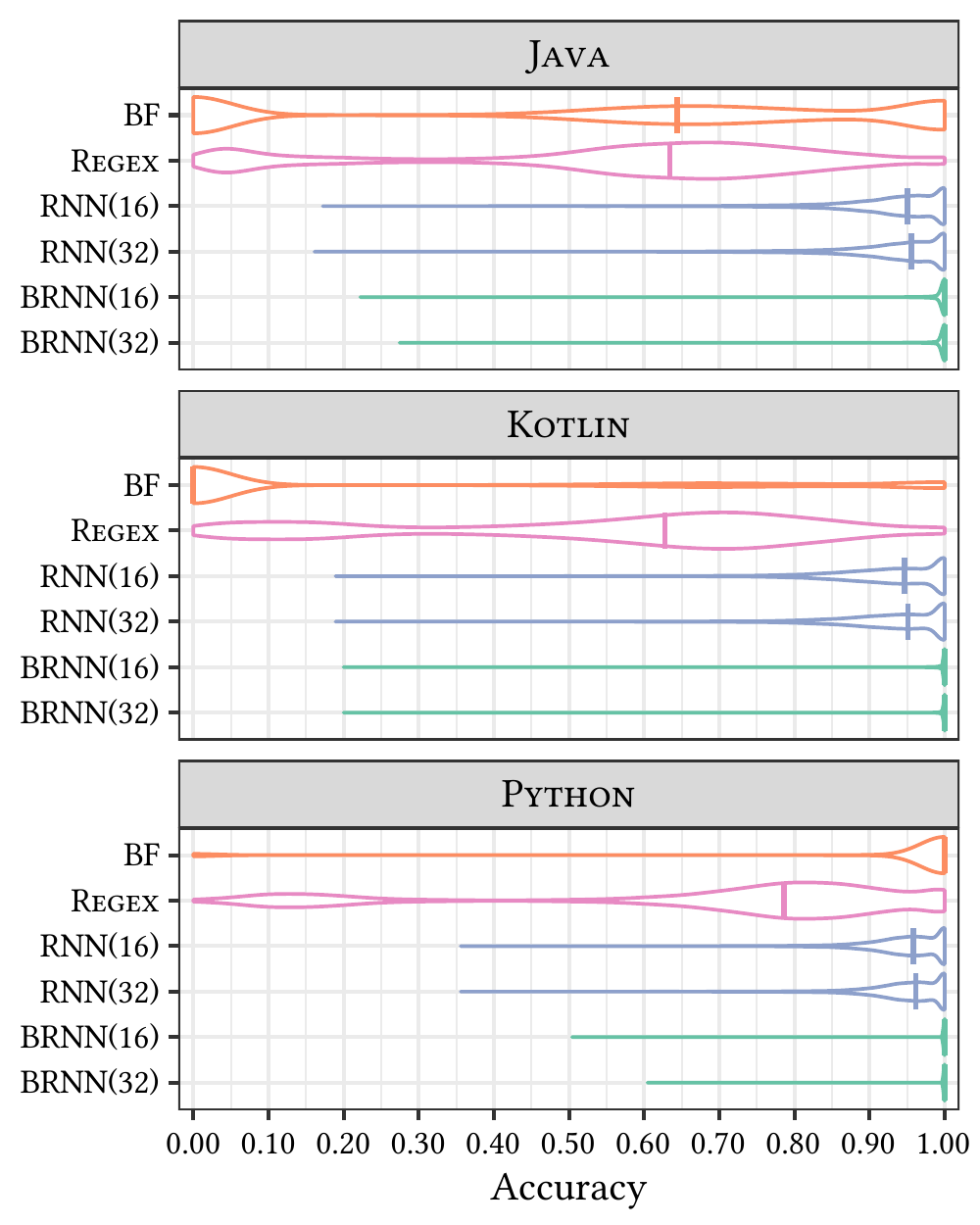}
    \caption{Accuracy values comparison for incomplete language derivations.}
    \label{fig:results:boxplot_rq4}
\end{figure}

\Cref{fig:results:boxplot_rq4} informs best about not only how consistently poorer the results of the \apprregex approach are compared to those of the \acp{rnn} and \acp{brnn}, but also how much more variable they can be expected to be.
Instead, the \ac{bf} resolvers proved to be the least eclectic strategy.
For \java, median performance values are close to those obtainable with the \apprregex resolver, however, at the cost of much greater variability than the latter.
\ac{bf} strategy performs the worst with \kotlin, yielding \num{0} median accuracy value and yet again a significant accuracy variance.
Finally, in the case of \python, the \ac{bf} approach is capable of outperforming both \apprregex and the base \ac{rnn} approaches, nearing the predictions of the \ac{brnn} models.
However, the latter presents a mildly smaller number of outliers.

% [TODO] Do we want to include this?
% Such behaviour is the result of the \python grammar, of which the native support for scripting means that line statements are less likely to require specific scaffoldings to be parsed, as it is the case for Java and Kotlin.

\section{Related Work}
\label{sec:related_work}

The main goal of this proposed approach is to show that deep learning can be used to perform syntax highlighting effectively and efficiently.
In the following, the current state-of-the-art approaches that most relate to the proposed approach are listed.

\paragraph{Deep learning type inference}
Similar applications of \ac{dl} models have been utilized in the field of \emph{Type Inference}; an example of this is \textsc{DeepTyper}~\cite{hellendoorn_deep_2018}.
In this case, motivated by the maintainability and readability benefits of a statically typed codebase, the model aims at aiding developers in the transition of code of dynamically typed languages supporting type annotation to their annotated equivalent.
Similar to how the proposed approach to \ac{sh} learns to infer the behavior of a parser on token ID sequences, \textsc{DeepTyper} aims at statistically inferring the compiler's type inference process.
Such capability becomes especially useful in languages such as \textsc{JavaScript}, which cannot deterministically handle \emph{duck-typing} even during runtime.
The architecture used in \textsc{DeepTyper} is also based on \acp{brnn}, however including extra infrastructure for the handling of more complex predictions.
In fact, this consists of bidirectional \ac{gru}~\cite{cho_learning_2014}, with \num{2} hidden layers of \num{650} hidden units each.
To proxy between the two hidden layers, an extra layer is introduced: the \emph{Consistency Layer}.
This pushes forward an extra input for the second \ac{brnn} layer, in the form of the average token representation (embeddings) of the first \ac{brnn} layer, thereby promoting the model to use long-range values in the input.
Furthermore, the model maps its input vector through an embedding layer of size \num{300}.
Finally, \textsc{DeepTyper} maps the values of its output layer through a \emph{softmax} function to obtain for each input token a categorical probability distribution over the types in some vocabulary.
The oracle is also generated analytically, with \textsc{TypeScript} files first annotated by the compiler and then stripped of their type annotations to obtain \textsc{JavaScript} files.

Unlike the approach proposed in this paper, \textsc{DeepTyper} uses tokens as inputs, complete of identifiers: this also allows it to compute type names.
However, this extra information is not needed in the \ac{sh} scope, in which structure is directly dependent on the sequence of token rule or type, \ie, \emph{tr}.
The adaptation of the \textsc{DeepTyper} model to the task of \ac{sh}, although obviously possible, is vain due to the evidence being reported.
Base \ac{brnn} models are never saturated in their ability to reach perfect \ac{sh} accuracy.
It means the extra infrastructure of a \textsc{DeepTyper} model would likely not generate better results but would lead to larger and slower models.

\paragraph{Learning lenient parsing and typing via indirect supervision}
\textsc{TypeFix} is a transformer~\cite{vaswani_attention_2017} decoder network developed as part of an approach to leniently parse and type \java code fragments~\cite{ahmed_learning_2021}.
It develops from the architecture and task of \textsc{DeepTyper}, and derives a deeper model based on a \num{6} layer decoder network, with each layer having multi-head attention and feed-forward.
By design, such flavors of encoder-decoder models promote the output of each inner layer to be a function of all combinations of units in the previous layer.
It promotes the learning of generalizable reductions of the relationships among elements in the input sequence.
More levels of relationships between the inputs may also be learned through \emph{multi-headed attention}, by adding more attention layers to the model.
Moreover, this mechanism allows the model to be more easily trained on long sequence, unlike \ac{rnn} models, which, due to their recurrent evaluation of an input vector, suffer from vanishing gradients~\cite{bengio_learning_1994}.
Similarly to the proposed strategy to \ac{sh}, and \textsc{DeepTyper}, \textsc{TypeFix} is trained over a synthetically derived oracle.
In particular, this consists of bindings of \java token identifiers and the respective deterministically derived type.
Hence, the model is trained to bind a categorical probability distribution over some fixed type vocabulary.

The reasons for which such architecture is not being evaluated in this first iteration towards on-the-fly \ac{sh}, are in line with those given for \textsc{DeepTyper}.

\paragraph{Generating robust parsers using island grammars}
Island grammars~\cite{moonen_generating_2001} are grammars which define both \emph{island} and \emph{water} productions.
\emph{island} rules define how to consume specific subsequences of some input sequence.
Instead, \emph{water} rules define how to consume all of those tokens that could not be bounded to any \emph{island} rule.
Such grammar structure might be used for the task of \ac{sh}.
In fact, given a language, one can define the set of \emph{island} rules as the collection of those sub-productions which consume highlightable sequences and map every other token to a particular production that consumes any terminal symbol.

Nevertheless, this strategy is outside the goals of this work.
Producing an \emph{island} grammar would induce a development workflow similar to the current state of practice, requiring developers to have a deep understanding of the grammatical structure and undertake a tedious process for the definition of productions with high coverage and accuracy.
It is significantly more challenging than providing a tree walker for relevant constructions of the original grammar, which by design correctly consumes all the valid iterations for the same feature.
Moreover, the \emph{island} approach would still leave the handing of incomplete language derivations in the hands of the developer.
%\comm{Pasquale}{I don't understand the following...}
Similarly to the state of practice, \emph{island}-like solutions represent the workflow this paper wishes to avoid.
%Today's state of practice, \ie, \ac{re}, and \emph{island}-like approaches are the very workflow from which the proposed strategy to \ac{sh} aimed at departing.

\section{Conclusions and Future Work}
\label{sec:conclusions}

The proposed approach is capable of consistently computing perfect \ac{sh} schemes for the average input files for all the mainstream languages considered.
Thereby, it comfortably outperforms the \ac{sh} accuracy achievable with the here tested state of practice.
Furthermore, this solution to \ac{sh} is capable of producing such outputs in expected time delays significantly faster and with lower variance than formal approaches, \ie, \acf{bf}, capable of equal outputs.
However, it is verified that for cases in which the language's grammar results in an efficient parsing of the input, as it is true for \python, the deep strategy does not represent a \begin{revised}superior alternative to the \apprbf with regards to the prediction delays,\end{revised} with both solutions yielding time delays suitable for these scenarios.

Future work might investigate further the accuracy with regard to the distribution of online snippets: an aspect that, due to the strict design of a \ac{bf} method, at this stage was not achievable.
For this purpose, the automated \textsc{APIzation} protocol of code fragments presented by Terragni and Salza~\cite{terragni_apization_2021}, might be used for the construction of grammatically correct versions of online snippets, from which a formal oracle could be derived.
Moreover, the native parallelisation of \acfp{cnn}~\cite{lecun_convolutional_1995}, already employed in sequence to sequence translation tasks~\cite{gehring_convolutional_2017}, may be exploited for the achieving of smaller prediction delays.

\section*{Acknowledgements}

The research leading to these results has received funding from the Swiss National Science Foundation (SNSF) project \enquote{Melise - Machine Learning Assisted Software Development} (SNSF204632).

% Bibliography
\balance
\bibliography{references, urls}

\end{document}